\documentclass[12pt,preprint]{aastex}

\slugcomment{To be resubmitted to Astrophysical Journal (revised
version)}

\shorttitle{Quantified wave transformations in the solar wind}
\shortauthors{Gogoberidze, Rogava \& Poedts}

\begin{document}

\title{Quantifying shear-induced wave transformations in the solar wind}

\author{Grigol Gogoberidze}
\affil{Georgian National Astrophysical Observatory, Kazbegi ave.
$2^a$ Tbilisi-0160, Georgia.} \email{gogober@geo.net.ge}

\author{Andria Rogava\altaffilmark{1}}
\affil{Centre for Plasma Astrophysics, K.U.Leuven, Celestijnenlaan
200B, 3001 Leuven, Belgium; Abdus Salam International Centre for
Theoretical Physics, Trieste I-34014, Italy.}
\email{Andria.Rogava@wis.kuleuven.be}

\author{Stefaan Poedts}
\affil{Centre for Plasma Astrophysics, K.U.Leuven, Celestijnenlaan
200B, 3001 Leuven, Belgium.}
\email{Stefaan.Poedts@wis.kuleuven.be}

\altaffiltext{1}{On leave from the Georgian National Astrophysical
Observatory, Kazbegi ave.\ $2^a$ Tbilisi-0160, Georgia.}

\begin{abstract}

The possibility of velocity shear-induced linear transformations of
different magnetohydrodynamic waves in the solar wind is studied
both analytically and numerically. A quantitative analysis of the
wave transformation processes for all possible plasma-$\beta$
regimes is performed. By applying the obtained criteria for
effective wave coupling to the solar wind parameters, we show that
velocity shear-induced linear transformations of Alfv\'en waves into
magneto-acoustic waves could effectively take place for the
relatively low-frequency Alfv\'en waves in the energy containing
interval. The obtained results are in a good qualitative agreement
with the observed features of density perturbations in the solar
wind.

\end{abstract}


\section{Introduction}

In spite of several decades of intense studies, the problem of the
connection between the physics of the inner parts of the solar
atmosphere and the solar wind remains incompletely understood. There
is a general consensus that the mechanisms of coronal heating are
somehow related to the mechanisms responsible for the acceleration
of the solar wind. The so-called ``basal" coronal heating (at $r <
1.5\;R_{\bigodot}$) is usually attributed to a mixture of processes,
e.g.\ wave dissipation (by phase mixing and/or resonant absorption),
magnetic reconnection, turbulence and plasma instabilities
\citep{cra02,cra04}. In `open' magnetic flux tubes which are feeding
the fast solar wind at $r > 2\;R_{\bigodot}$ where the solar plasma
is largely collision-less, however, additional heating seems to take
place. This conclusion follows from a number of \emph{bona fide}
observational signatures, viz.\ (a)~low electron temperatures
($T_e<1.5 \times 10^6 \;{\rm K}$) in coronal holes, (b)~\emph{in
situ} observations of the temperature anisotropy $T_p
> T_e$ at 1~AU, (c)~low radial gradients of these temperatures
throughout the solar wind. These signatures unequivocally indicate
at a gradual, temporally and spatially extended addition of energy
\citep{cb03} to the solar wind (the so-called ``extended coronal
heating") at a wide range of heliocentric distances.

Presumably, propagating perturbations -- magnetohydrodynamic (MHD)
waves, vortices, shocks, turbulent eddies -- are the best
candidates for the transmission of energy from the inner to the
outer parts of the solar atmosphere and further into the solar
wind. In particular, Alfv\'en waves are believed to play a
significant role in the coronal heating and the subsequent
acceleration of the solar wind \citep{cb05}. They are also
considered as important diagnostic tools of the various physical
processes occurring in solar plasmas.

The presence of these efficient ``energy-transmitters" is currently
quite convincingly established. Even though previous observational
studies only emphasized the appearance of Alfv\'en waves in the
solar atmosphere, in the modern era of satellite-based solar studies
it became increasingly clear that the solar atmosphere hosts all
three basic MHD wave modes, i.e.\ Alfv\'en waves and both slow and
fast magneto-acoustic waves. Many different sets of observations
contributed to this ever growing evidence: in coronal loops the
evidence came from the Extreme Ultra Violet (EUV) data from the
TRACE satellite; in coronal plumes (slow magneto-acoustic waves)
from the EIT instrument on-board SOHO; in the solar wind
(propagating Alfv\'en waves) from \emph{in-situ} Helios and Ulysses
spacecraft measurements \citep{cra02}. Therefore, both the
observational data and the current theoretical studies are strongly
in favor of the presence of a causal bond between the heating of the
solar corona and the acceleration of the solar wind and some,
probably significant, role of the MHD waves in the functioning of
this bond. However, it is still unknown up to what extent the fluxes
of the solar wind's mass, momentum, and energy are driven by
``classic", Parker-type gas pressure gradients, and what is the
relative share of wave-flow (wave-particle) interactions and/or
turbulence \citep{cra04}.

It is widely believed that one of the main mechanisms responsible
for the heating of the solar wind, is the turbulent cascade of
Alfv\'enic fluctuations. The advantage of this heating model is that
it can also explain the above-mentioned relatively high proton
temperatures at 1~AU \citep{grm95}. Remarkably, it was also shown
\citep{gr99} that the turbulent cascade in the solar wind seems to
evolve most rapidly in areas with a formidable velocity shear. At
the same time, the measurements of significant density fluctuations
($\delta\rho/\rho\sim 0.1$) in the solar wind \citep{grm95} are
usually interpreted as evidence for the presence of magneto-acoustic
waves not only in the corona but also in the solar wind. On the
other hand, both slow and fast magneto-acoustic modes are expected
to be strongly damped by Landau damping for solar wind parameter
values \citep{bar79}. Therefore, there is a very scarce, if any,
chance for their undamped propagation from the solar corona up to
the outer wind regions. This logically leads to the following
presumable solution of this puzzle: the magneto-acoustic waves in
the solar wind should have a local origin.

It has been suggested that three- and/or four-wave resonant
processes may be of a considerable importance in the solar wind
\citep{lm80,bn01,c05}. However, after studying the nonlinear
interaction of oblique fast magneto-acoustic waves with Alfv\'en
waves in the solar wind, Lacombe and Mangeney came to the conclusion
that ``this non-linear process is not very efficient in the solar
wind, so that Alfv\'en waves can be considered as decoupled from
compressive waves in the major part of the m.h.d. spectral range"
\citep[see][Abstract]{lm80}. Another argument against this scenario
is that, if one of the modes is strongly damped, three- and/or
four-wave resonant processes transform into the induced scattering
of the involved waves by plasma particles \citep{bzm73} and, hence,
they do not lead to the reappearance of the damped wave mode. In
addition, there are observational arguments against the multi-wave
scenario. It is known \citep{bis03} that in the slow solar wind the
fluctuation amplitudes are smaller then in the fast solar wind. If
nonlinear multi-wave resonant processes were responsible for the
generation of compressive fluctuations, then one should expect a
much higher level of density fluctuations in the fast wind than in
the slow wind. However, observations show that the slow solar wind
is usually much more compressional then the fast solar wind
\citep{bb91,bpb04}. Consequently, the local generation of
magneto-acoustic waves could hardly be explained by multi-wave
resonant processes.

Hence, the gradual appearance of locally-produced magneto-acoustic
waves throughout the solar wind seems to be an observationally
confirmed fact, but still solicits for an adequate explanation. In
the present paper, we argue that the velocity-shear-induced linear
mode conversion of Alfv\'en waves into fast and slow
magneto-acoustic waves represents an efficient mechanism for the
generation of the observed magneto-acoustic waves within the solar
wind. It is also argued that the velocity shear could have an
important contribution in the gradual ``energization" of the solar
wind via the generation of fast and slow magneto-acoustic MHD waves,
that are eventually strongly damped by Landau damping. Our analysis
shows that the linear mode conversion of Alfv\'en waves into
magneto-acoustic waves is especially efficient for the low-frequency
Alfv\'en waves from the energy containing interval. Afterwards,
these perturbations can cascade to the smaller scales \citep{mbm87}
by turbulence.

It is well-known \citep{rgmg92,gr99} that velocity shear is one of
the most important ingredients in the solar wind dynamics. Recent
studies, based on in situ observations, showed that the fast/slow
wind interface in the interplanetary space has two parts: a smoothly
varying ``boundary layer" flow that separates the fast wind from the
coronal holes, and a sharper discontinuity between the slow and the
intermediate solar wind \citep{s05}. A relatively high velocity
shear was observed by \emph{Ulysses} over its first orbit in the
transition area between the fast and slow solar winds at
$13^\circ-20^\circ$ latitudes. The analysis of the data
\citep{mrgbf98} showed that the transition area between the two
winds consisted of two regions, the first one with a width $\Delta
l_1\approx 2~10^7~{\rm km}$ and a velocity change $\Delta V_1
\approx 200~{\rm km \, s^{-1}}$, and the second one with $\Delta
l_2\approx 8~10^7~ {\rm km}$ and $\Delta V_2 \approx 100~{\rm km \,
s^{-1}}$. However, during \emph{Ulysses'} second orbit, the global
solar wind structure was remarkably different from that observed
during its first orbit \citep{m01,m03}. Overall, the solar wind was
the slowest seen thus far in the satellite's ten-year journey (~270
km s$^{-1}$). The wind was highly irregular, with less pronounced
periodic stream interaction regions, more frequent coronal mass
ejections, and only a single, short interval of fast solar wind. The
complicated solar wind structure obviously was related with a higher
complexity of the solar corona around the solar activity maximum,
e.g.\ with the disappearance of large polar coronal holes and with
the presence of smaller-scale coronal holes, frequent CMEs and
coronal streamers.

Originally, the idea of (velocity-)Shear-induced Wave
Transformations (SWTs) in the solar wind was proposed by
\citet{prm98}. However, the model of these authors was based on a
phase-space analysis of the temporal evolution of individual
fluctuation harmonics in the ideal (viscosity- and resistivity-less)
limits, and it was only of a qualitative nature. Since then,
considerable progress has been made in three important directions:
(i)~real (physical) space numerical simulations have been performed
and it has been shown that SWTs occur in a well-pronounced way and
they lead to easily recognizable collective phenomena in MHD plasma
flows \citep{bprr01}; (ii)~the role of the dissipation has been
analyzed and the concept of shear-induced \emph{self-heating} was
introduced \citep{rog04} and it was found that compressible wave
modes (e.g., sound waves in hydrodynamic flows and fast
magneto-acoustic waves in MHD flows) grow non-exponentially and
undergo subsequent viscous and/or resistive damping leading to the
heating of the ambient flow by ``inborn" waves; (iii)~a noteworthy
quantum-mechanical (QM) analogy has been disclosed that helped to
apply efficient mathematical tools from the scattering matrix theory
of QM to the SWT studies which helped to give a quantitative rigor
to this theory and to calculate directly the efficiency of different
wave transformation channels by determining the corresponding
transformation coefficients.

The aim of the present paper is to apply the latter method to the
study of MHD wave transformations in the solar wind and to give a
fully quantified description of the coupling efficiency by
calculating the corresponding transformation coefficients. The
mathematical methods used in this paper are similar to the ones that
originally were developed in the 1930s for quantum mechanical
problems \citep{stu32,zen32,lan32}. More recently, similar
asymptotic methods have been successfully applied to various other
problems including the interaction of plasma waves in inhomogeneous
media \citep{swa98,gcsl04,RG05}.

The present paper has the following structure: the main mathematical
consideration is presented in the next section. Different regimes of
wave transformations, depending on the value of the plasma-$\beta$,
are studied in the third section. A brief discussion and the
conclusions are given in the final section.

\section{Basic Formalism}

In this section, we give a brief synopsis of the basic shear flow
model that was studied by \citet{prm98}. In this vein, we consider a
plane-parallel, compressible, magnetized, unbounded shear flow with
uniform values of the equilibrium plasma density $(\rho_0)$ and
pressure $(P_0)$. The equilibrium magnetic field $\bf B_0$ is
considered to be uniform as well and is assumed to be directed
parallel to the flow velocity which, in turn, is spatially
inhomogeneous and oriented in the $z$-direction:
$$
{\bf U}_0=(0,0,A x), \eqno(1)
$$
with the shear parameter $A>0$ being defined as a positive constant.

The linearized ideal MHD equations governing the evolution of the
perturbations of the plasma density $(\rho^{\prime})$, pressure
$(p^{\prime})$, velocity field $(\bf{u^{\prime}})$ and magnetic
field $(\bf {b^{\prime}})$ in this flow are [with the notation
${\cal D}_t \equiv
\partial_t + {\bf U}_0 \cdot {\bf \nabla}$]:
$$
{\cal D}_t \rho^{\prime} + {\rho_0} {\bf \nabla} \cdot
{\bf{u^{\prime}}} = 0, \eqno(2)
$$
$$
{\rho_0}[{\cal D}_t{\bf{u^{\prime}}} + ({\bf u^{\prime}} \cdot
{\bf \nabla}) {\bf U}_0] =- {\bf \nabla} p^{\prime} -
\frac{1}{4\pi} {\bf B}_0 \times ({\bf \nabla} \times
{\bf{b^{\prime}}}),  \eqno(3)
$$
$$
{\cal D}_t{\bf{b^{\prime}}} - ({\bf B}_0 \cdot {\bf \nabla}
){\bf{u^{\prime}}} + {\bf B}_0 ({\bf \nabla} \cdot
{\bf{u^{\prime}}})= 0, \eqno(4)
$$
$$
{\bf \nabla} \cdot {\bf{b^{\prime}}} = 0. \eqno(5)
$$

The standard technique of the so-called ``shearing-sheet
approximation" \citep{gl65} implies that it is convenient to expand
the perturbations as follows:
$$
\mbox{\boldmath $\Phi$}^\prime({\bf x},t) = \mbox{\boldmath
$\Phi$}^\prime ({\bf k},t) \exp \left[ i(k_x(t)x + k_yy+k_zz)
\right], \eqno(6)
$$
where the state vector $\mbox{\boldmath $\Phi$}^\prime \equiv [{\bf
u}^{\prime}, {\bf b}^{\prime}, \rho^\prime, p^\prime]$, while
$k_x(t) \equiv k_x(0) - k_z A t$, and $k_x(0)$, $k_y$, and $k_z$ are
the initial ($t=0$) values of the components of the wave number
vector ${\bf k}(t)$. Remarkably, this ansatz reduces the
mathematical part of the task to the solution of a closed set of
ordinary differential equations (ODEs) in time: a solvable initial
value problem. Note that $k_x(t)$ varies in time and this fact is
usually referred to as the ``drift" of the Spatial Fourier Harmonics
(SFH) in the phase ${\bf k}$-space \citep{crt96,prm98}.

From the set of Eqs.~(2)--(5), considering adiabatic perturbations
$p^\prime = {c_s}^2 \rho^\prime$, where $c_s$ is the sound speed,
and introducing the following non-dimensional parameters and
variables $R \equiv A/(V_A k_z)$, $\tau \equiv k_z V_At$, $\beta
\equiv c_s^2/V_A^2$, $K_y \equiv k_y/k_z$, $K_x(\tau) \equiv
k_x/k_z-R \tau$, $K^2(\tau)=K_x^2(\tau)+K_y^2 + 1$, $\rho({\bf
k},\tau)\equiv i {\rho^{\prime}({\bf k},\tau)}/{\rho_0}$, ${\bf
b}({\bf k},\tau) \equiv i{{\bf b}^{\prime}({\bf k},\tau)}/{B_0}$,
${\bf v}({\bf k}, \tau) \equiv {\bf u}^{\prime}({\bf k},\tau)/V_A$,
$\psi \equiv \rho + K_x(\tau)b_x + K_yb_y =\rho - b_z$, ($V_A \equiv
B_0/ \sqrt{4 \pi \rho_0}$ is the Alfv\'en speed), one can derive the
following set of ODEs \citep{prm98} [$F^{(n)}\equiv\partial^n_tF$]:
$$
\psi^{(2)} + {\cal C}_{11}\psi + {\cal C}_{12}(\tau)b_x+{\cal
C}_{13}b_y = 0, \eqno(7)
$$
$$
b^{(2)}_x + {\cal C}_{22} (\tau) b_x +{\cal
C}_{21}(\tau)\psi+{\cal C}_{23}(\tau)b_y=0, \eqno(8)
$$
$$
b^{(2)}_y + {\cal C}_{33} b_y + {\cal C}_{31}\psi+{\cal
C}_{32}(\tau)b_x=0, \eqno(9)
$$
where $||{\cal C}||$ is a $3 \times 3$ symmetric matrix defined
as:
$$
||{\cal C}|| = {\left(\matrix{\beta& - \beta K_x(\tau) & -\beta
K_y \cr - \beta K_x(\tau) & 1+(1+\beta)K_x^2(\tau) & (1+\beta)
K_x(\tau) K_y \cr - \beta K_y  & (1+\beta) K_x(\tau) K_y  &
1+(1+\beta)K_y^2 \cr}\right)}.\eqno(10)
$$

The total energy of the perturbations, in the non-dimensional
form, can be defined as the sum of the kinetic, the magnetic and
the compressional energies. It can be written down in the
following way:
$$
E({\bf k},\tau) \equiv {1 \over 2}{\biggl[|\textbf{v}|^2 +
|\textbf{b}|^2+ \beta \rho^2\biggr]}. \eqno(11)
$$

From Eqs.~(7)--(9) it is easy to see that in the absence of the
shear in the flow (i.e.\ for $R=0$, and thus $A=0$), these equations
describe independent oscillations with the fundamental
eigenfrequencies \citep{prm98}:
$$
{\Omega^2_{s,f}}=\frac{1}{2}(1+\beta) K^2 \left[1 \pm
\sqrt{1-\frac{4\beta}{(1+\beta)^2 K^2}} ~\right],  \eqno(12a)
$$
$$
{\Omega^2_A}=1, \eqno(12b)
$$
that can be easily identified as fast and slow magneto-acoustic
waves (FMW and SMW) and Alfv\'en waves (AW), respectively. For the
eigenfunctions, $\Psi_i$, corresponding to the above-defined
$\Omega_i$ eigenvalues, we find \citep{gcsl04}:
$$
\Psi_f = \frac{(\Omega_S^2 - \beta)\rho + \Omega_S^2 (K_x b_x +
K_y b_y )}{\sqrt{(\Omega_S^2 - \beta)^2+\beta^2 K_\perp^2}}, \eqno(13a)
$$
$$
\Psi_s = \frac{\beta {K_\perp^2} \rho + (\Omega_S^2 - \beta K^2
)(K_x b_x + K_y b_y )}{K_\perp \sqrt{(\Omega_S^2 -
\beta)^2+\beta^2 K_\perp^2}}, \eqno(13b)
$$
$$
\Psi_A = \frac{K_y b_x - K_x b_y}{\sqrt{K_\perp^2}}. \eqno(13c)
$$
where $ K_\perp^2 \equiv K_y^2+K_x^2$.

Notice than, when the flows are only weakly sheared (i.e.\ when $R
\ll 1$), the coefficients in Eqs.~(7)--(9) vary only slowly or
adiabatically. This implies that the expressions for the fundamental
eigenfrequencies, given by Eq.~(12), and the corresponding
eigenfunctions, given by Eq.~(13), are still useful for a
qualitative description of the shear-induced dynamics of the wave
modes and their coupling/conversion properties \citep{crt96,prm98}.

\section{Transformation regimes: the transition matrix method}

As will be shown in the next section, the characteristic values of
the normalized velocity shear rate in the solar wind plasma always
satisfy the condition $R \ll 1$. This necessarily implies that the
coefficients in Eqs.~(7)--(9) are only slowly varying functions of
$\tau$ and, therefore, the adiabatic (WKB) approximation holds
everywhere except in the immediate vicinity of the turning
[$\Omega_i(\tau_t)=0$] and the resonant
[$\Omega_1(\tau_r)=\Omega_2(\tau_r)$] points. Using Eq.~(12) one can
evaluate that the condition:
$$
\dot{\Omega}_i \ll {\Omega}_i^2, \eqno(14)
$$
is satisfied for all the MHD wave modes at any moment of time, or
equivalently, none of the turning points are located near the real
$\tau$-axis. From a physical point of view, this means that there
are no (over-)reflection phenomena \citep{gcsl04} and resonant
coupling can only occur between different waves modes with the same
sign of the phase velocity.

A careful analysis of the system yields that the resonant coupling
takes place in the vicinity of the point $\tau_\ast$ where
$K_x(\tau_\ast)=0$ \citep{prm98}. According to the general theory of
such systems, the timescale of the resonant coupling $\Delta \tau$
is of the order of $\Delta \tau \sim R^{-n/(n+1)}$ \citep{gcsl04},
where $n$ indicates the order of the resonant point \footnote{The
resonant point is said to be of the order $n$ when
$(\Omega_1^2-\Omega_2^2) \sim (\tau - \tau_{r})^{n/2}$ in the
neighborhood of $\tau_{r}$}, and, therefore, the evolution of the
waves is adiabatic when
$$
|K_x(\tau)| \gg R^{1/(n+1)}. \eqno(15)
$$

If this condition is satisfied, the temporal evolution of the waves
is described by the standard WKB solutions:
$$
\Psi_i^\pm = \frac{D_i^\pm}{\sqrt {\Omega_i(\tau)}}e^{\pm i \int
\Omega_i(\tau)d\tau}, \eqno(16)
$$
where the $D_i^\pm$ denote the WKB amplitudes of the wave modes with
positive and negative phase velocity along the $z$-axis,
respectively. All the physical quantities can be readily found by
combining the Eqs.~(7)--(9). The energies of the involved wave modes
satisfy the standard adiabatic evolution condition \citep{prm98}:
$$
E_i = \Omega_i(\tau) (|{D_i^+}|^2 + |{D_i^-}|^2). \eqno(17)
$$
From this equation it follows that $|{D_i^\pm}|^2$ can be
interpreted as the number of `wave particles' (the so-called
`plasmons'), in analogy with quantum mechanics.

Let us assume that initially $K_x(0) \gg R^{1/(n+1)}$. Due to the
linear drift in the ${\bf k}$-space, $K_x(\tau)$ decreases and when
the mode enters the ``degeneracy area" \citep{prm98}, the mode
dynamics becomes non-adiabatic due to the resonant coupling between
the modes. Afterwards, when $K_x(\tau) \ll-R^{1/(n+1)}$, the
evolution becomes adiabatic again. Denoting the WKB amplitudes of
the wave modes before and after the coupling region by
${D_{i,B}^\pm}$ and ${D_{i,A}^\pm}$, respectively, and by employing
the formal analogy with the $S$-matrix of the scattering theory
\citep{kop95}, one can connect ${D_{i,A}^\pm}$ with ${D_{i,B}^\pm}$
via the so-called transition matrix:
$$
{\bf D}^+_{A} = {\bf T} {\bf D}^+_{B}, ~~~{\bf D}^-_{A} = {\bf
T}^{\ast} {\bf D}^-_{B}, \eqno(18)
$$
where ${\bf T}$ and its Hermitian conjugated matrix ${\bf T}^\ast$
are $3\times3$ matrices. Note, that none of the turning points are
located near the real $\tau$-axis and, therefore, there is no
transition between the modes with opposite signs of the phase
velocity along the $z$-axis \citep{gcsl04}.

Notice that all the coefficients in the governing equations are
real. Moreover, the matrix given by Eq.~(10) is symmetric. As a
consequence \citep{kop95,fed83}, the transition matrix ${\bf T}$ is
unitary \footnote{A matrix ${\bf U}$ is {\it unitary} if its
conjugate transpose, ${\bf U}^{H}$ is equal to the inverse matrix
${\bf U}^{-1}$.}, and
$$
\sum_j |T_{ij}|^2 =1. \eqno(19)
$$
Generally speaking, this equation represents the conservation of the
wave action. When $R \ll 1$, it transcribes  into the energy
conservation throughout the resonant coupling of the wave modes
\citep{gcsl04}. Energetically this means that a transformed wave
mode is generated solely on the expense of the energy of the
incident wave mode.

The crucial physical importance of this matrix follows from the fact
that the value of the quantity $|T_{ij}|^2$ represents a part of the
energy transformed via the resonant coupling of the modes. That is
why the absolute values of the transition matrix components
$|T_{ij}|$ are called the `transformation coefficients' of the
corresponding wave modes. For the resonant interaction of two wave
modes, e.g.\ $i$ and $j$, the unitarity of the matrix ${\bf T}$
guarantees an important property of the transition matrix, viz.\
$T_{ij}=T_{ji}$. This symmetry property holds for the resonant
interaction of two wave modes only. If, in the same time interval,
there is an effective coupling of more then two wave modes, then the
symmetry property fails.

In earlier studies, a lot of attention was usually paid to the
resonant points of the first order \citep{ll77,fed83}. In this case,
only the dispersion equations of the waves are needed to derive the
transformation coefficients with accuracy $O(R^{1/2})$. For the
second and/or higher order resonant points, analytical expressions
for transformation coefficients can be derived only in the case of
weak interactions ($ T_{ij}\ll 1,~~ i \neq j$). As a matter of fact,
if the resonant points are not close to the real $\tau$ axis, in the
sense that
$$
\phi_{ij} = \left| {\rm Im}
\int_{\tau_0}^{\tau_{r}}(\Omega_i-\Omega_j)d\tau \right|\gg 1,
\eqno(20)
$$
then the transformation coefficient is just equal to:
$$
T_{ij} \approx \frac{\pi}{2} \exp\left(- \phi_{ij} \right).
\eqno(21)
$$
Here, and hereafter, the signs of the absolute magnitude for
transformation coefficients are omitted, i.e., from now on the
notation $T_{ij}$  means $|T_{ij}|$.

In the context of the {\em linear} problem that we are studying in
this paper, viz.\ the velocity-shear-induced coupling of MHD waves,
all the resonant points are of the second order, as will be shown
later.

In the following subsections, we will study in detail the coupling
of MHD wave modes (AW, FMW, and SMW) for different regimes of the
plasma-$\beta$ that are interesting in the solar wind context. For
the earlier, detailed, but only qualitative analysis of the same
regimes see \citet{RPM00}.

\subsection{The case $\beta \ll 1$}

It is well-known that the magnetic field dominates the plasma
($\beta \ll 1$) throughout most of the solar corona, especially at
lower altitudes and within the so-called `active' regions, as well
as in the innermost regions of the solar wind. For the MHD waves in
this regime, the frequency of the SMW is far smaller then the
frequencies of the FMW and the AW. Therefore, the coupling of the
SMW with the other two MHD modes is exponentially small with respect
to the large parameter $1/R$ and can be neglected. Consequently, in
the set of the governing equations Eqs.~(7)--(9), the equation for
the variable $\psi$ decouples from the other equations, and the
equations for $b_x $ and $b_y$ describe the evolution of the coupled
AW and FMW:
$$
{\ddot b}_x  + \left[ 1+ K_x^2(\tau) \right] b_x = -K_x(\tau) K_y
b_y, \eqno(22)
$$
$$
{\ddot b}_y + \left[ 1 + K_y^2 \right] b_y = -K_x(\tau) K_y b_x.
\eqno(23)
$$

The normalized frequencies of the coupled wave modes are:
$\Omega_f^2(\tau) = 1 + K_y^2 + K_x^2(\tau),~\Omega_A^2 = 1$.
Therefore, there are two, complex conjugated, second order
resonant points:
$$
K_x(\tau_{fA}) = i K_y,~~~K_x(\tau_{fA}^\ast) = -i
K_y. \eqno(24)
$$

The condition (15) then implies that the evolution of the waves is
adiabatic if
$$
\delta \equiv |K_y^3|/R \gg 1, \eqno(25)
$$
and if this condition is satisfied, Eq.~(21) yields:
$$
T_{fA} = \frac{\pi}{2} \exp\left( -\frac{\delta^3}{3} \right).
\eqno(26)
$$

An analytical expression for the transformation coefficients can
also be derived in the opposite limit, $\delta \ll 1$. In this
case \citep{gcsl04}:
$$
T_{fA} \approx \frac{2^{2/3}\pi}{3^{1/3}\Gamma \left(
\frac{1}{3}\right)} \delta \left( 1 - \frac{\Gamma\left(
\frac{1}{3}\right)}{2^{7/4} 3^{1/3} \Gamma \left(
\frac{2}{3}\right)} \delta^4 \right). \eqno(27)
$$

Results of the numerical solution of the initial set of equations
(7)--(9) (solid line) as well as of the analytical expressions given
by Eqs.~(26) (dash-dotted line) and (27) (dashed line) are presented
in Fig.~1. This plot shows that the transformation coefficient
attains its maximal value $(T_{fA}^2)_{max} = 1/2$ at $\delta^{cr}
\approx 0.89$.

In the case of first-order resonant points, the transformation
coefficient monotonically tends to unity when the resonant points
tend to the real axis \citep{ll77,fed83} and, therefore, the total
conversion of one wave mode into another wave mode is possible.
Here, in the case of second-order resonant points, existing for the
FMW-AW coupling in a low-$\beta$ plasma, $({T_{fA}}^2)_{max} = 1/2$.
This leads to the following important astrophysical conclusion: in
those regions of the solar wind where $\beta \ll 1$, even under the
most favorable conditions, at most half of the energy of the AWs can
be transformed into FMWs and vice versa!

Yet another important aspect of the coupling in this case is that,
if the resonant point tends to the real $\tau$ axis (i.e., $K_y
\rightarrow 0$), the transformation coefficient tends to zero.

\subsection{The case $\beta \gg 1$}

In the high-$\beta$ case, $\Omega_s,\Omega_A \ll \Omega_f$.
Therefore, only the coupling between the AW and the SMW can be
important in this regime. From the expressions for the fundamental
frequencies [see Eq.~(12)] one can easily deduce that, for the
AW-SMW coupling, there are two complex conjugated, second-order
resonant points, that are given by Eq.~(24). The condition (20) in
this case has the following form:
$$
\frac{1}{\beta R}\left| K_y -\frac {\mbox{arcsinh} \left( K_y
\right) }{\sqrt {1+K_y^2}} \right| \gg 1. \eqno(28)
$$

If this condition is satisfied, and in addition $K_y \ll 1$, we can
derive from Eq.~(21) the following formula for the transformation
coefficient:
$$
T_{sA} \approx \frac{\pi}{2} \exp\left( -\frac{|K_y|^3}{3\beta R}
\right). \eqno(29)
$$

Unlike in the low-$\beta$ case, now there is no unique parameter,
like the parameter $\delta$, which would give a complete description
of the transformation process. Instead, in this limit, there are two
``governing" parameters, viz.\ $K_y$ and $\beta R$.

Let us first consider the case $ \beta R \ll 1$. In this case,
Eq.~(28) reduces to $\delta_1 \equiv |K_y|/(\beta R)^{1/3} \gg 1$,
and the properties of the wave transformation are essentially the
same as in the case of the transformation of the AW to the FMW and
vice versa. As a matter of fact, if $\delta_1 \ll 1$, then the
leading term of the asymptotic expressions of the transformation
coefficient is given by Eq.~(27), with $\delta$ replaced by
$\delta_1$. At $ \beta R\ll 1$, the transformation coefficient
reaches its maximum $(T_{sA}^2)_{max} =1/2$ at $\delta_1^{cr}$ that
coincides with $\delta^{cr}$.

The dependence of the transformation coefficient on $K_y$, on the
other hand, provides some very interesting and new details, compared
with the low-$\beta$ case. On Fig.~2, we display the numerical
solution of the initial set of equations (7)--(9) for the cases
$\beta R = 0.025 $ and $\beta R = 1 $. This numerical inspection
shows a remarkable fact: when $\beta R$ is not small the properties
of the transformation process are quite different (see Fig.~2). As a
matter of fact, when $K_y \ll 1$, it turns out that the
transformation coefficient $T_{sA}$ does not depend on $\beta R $ at
all and is given by the formula $T_{sA} \approx 2.05 K_y$. Besides,
as it can be seen clearly from Fig.~2, $(T_{sA})_{max}=1$, i.e.\
now, unlike the low-$\beta$ situation, a total transformation of one
wave mode into another wave mode is possible!

\subsection{The $\beta \sim 1$ case}

The $\beta \sim 1$ case is the most interesting but also the most
complicated case of velocity-shear-induced MHD wave transformations.
The frequencies of all three modes are of the same order in this
case and a simplification or reduction of the set of Eqs.~(7)--(9)
is not possible. However, the analysis of Eq.~(12) enables us to see
that there is a pair of complex conjugated first-order resonant
points:
$$
{K_x}(\tau_{1,2}) = \pm i \sqrt{ K_y^2+\left(
\frac{\beta-1}{\beta+1} \right)^2}, \eqno(30)
$$
and another pair of complex conjugated resonant points of the second
order:
$$
{K_x}(\tau_{3,4}) = \pm i K_y.  \eqno(31)
$$

The essential complexity of this case stems from the fact that the
transition matrix method does not allow one to derive analytic
expressions for the transformation coefficients when more than two
wave modes are effectively coupled. However, the numerical study of
the problem shows (see for details \citet{gcsl04}) that the
qualitative character of the wave transformation processes in this
case is mainly similar to the cases described in the previous
subsections.

\section{Discussion and conclusions}

Several years ago, \citet{prm98} argued that velocity-shear-induced
MHD wave transformations could be an important ingredient of the
wave dynamics in the solar wind, contributing to its acceleration
and heating. It was further argued \citep{prm99,RPM00} that SWTs can
contribute to the transmission of the waves from the chromosphere,
through the transition layer to the corona and further into the
solar wind. Besides, it was speculated that ``self-heating" of the
solar plasma flows \citep{rog04,srp04,spp06}, via the agency of
mutually coupled wave modes, might be one of the reasons for the
coronal heating (both ``basal" and ``extended") and the acceleration
of the solar wind.

Up to now, the scenario of the linear coupling and the mutual
transformations of the MHD wave modes in the solar wind was outlined
only qualitatively. Clearly, this merely qualitative picture lacked
quantitative strength and rigor, since it was not clear how
efficient the SWTs could be in the solar wind context. In this
paper, we solved this problem and we provided a systematic,
quantitative description of SWTs in terms of the recently developed
transition matrix method (originally developed in the 1930s for
quantum mechanical applications) and transformation coefficients.

In particular, we have demonstrated that the dynamics of the MHD
wave conversions is determined by three key parameters: the flow
velocity shearing rate $R$, the plasma $\beta$ and the ratio
$K_y=k_y/k_z$. Since the aim of this paper is to verify whether the
SWTs are well-pronounced in the solar wind, it is necessary to
analyze whether these parameters within the solar wind really have
values which favor efficient mode transformation processes or not.

For the dimensionless shearing parameter $R$ we have
$$
R=\frac{\Delta V}{\Delta l}\frac{1}{\Omega_a}. \eqno(32)
$$
Using for $\Delta V$ and $\Delta l$ the values presented in the
Introduction, we found that even for very low frequency Alfv\'en
waves (with $\Omega_a > 10^{-4}~{\rm s^{-1}}$), the dimensionless
shear parameter satisfies the condition $R \ll 1$, for both regions
of the transition area between the fast and slow solar winds.
Bearing also in mind that the plasma $\beta \sim 1$ in the solar
wind, we obtain that $R \beta \ll 1$, and according to the
considerations in the previous section, the coupling of the AWs with
both the fast and slow magneto-acoustic waves has the same
qualitative character. As a matter of fact, Eq.~(25) gives us the
necessary condition for an effective coupling
$$
K_y \lesssim R^{1/3}, \eqno(33)
$$
showing that an effective coupling between the Alfv\'en mode and the
fast and slow magneto-acoustic modes takes place for the low
frequency Alfv\'en waves from the energy containing interval.
Indeed, using for the frequency of the AWs, $\Omega_a=2\pi/T_a
\approx 6\times10^{-4}~{\rm s^{-1}}$, and using for $\Delta V$ and
$\Delta l$ the values presented in the Introduction, we found that
$R_1\approx 1.6\times10^{-2}$ and $R_2\approx 2\times10^{-3}$ for
the two regions of transition between the fast and slow solar winds.
Equation~(33) then yields for the perpendicular wave numbers of the
waves, $k_y \lesssim 0.25 k_z$ for the first transition region and
$k_y \lesssim 0.13 k_z$ for second transition region, respectively.
Noting that in the energy containing range, the turbulent
fluctuations are nearly isotropic \citep{bbgv06}, we can expect that
for the typical turbulent fluctuations in this range parallel and
perpendicular wave numbers of the same order ($k_y \sim k_z$).
Consequently, the linear transformations of Alfv\'en waves to fast
and slow magneto-acoustic waves are quite likely to take place in
this frequency band.

On the other hand, the linear transformation of Alfv\'en waves to
magneto-acoustic waves seems to be inefficient in the high frequency
band, including the inertial interval of solar wind fluctuations. As
a matter of fact, Eqs.~(32) and (33) yield that the modes which
could be effectively transformed should satisfy $k_y \ll k_z$.
However, theoretical research of MHD turbulence as well as numerical
simulations and analysis of the solar wind data
\citep{smm83,gs95,cv00,mbg03,bis03,om05,g06} all indicate that the
energy cascade in the inertial interval proceeds much more
effectively in the direction perpendicular to the mean magnetic
field and, therefore, for high-frequency modes one usually has $k_y
\gg k_z$. Consequently, the condition (33) could be hardly fulfilled
for typical parameters of high-frequency Alfv\'en waves in the solar
wind.

Although the performed analysis allows us to make quite definite
predictions about the efficiency of linear conversion of AWs into
FMWs and SMWs, direct quantitative comparison with the observations
is difficult since the compressive part of the velocity field can
not be isolated by single spacecraft observations \citep{bis03}.
Consequently, it is difficult to associate the observed density
fluctuations to specific mode characteristics, such as fast or slow
magneto-acoustic modes.

Nevertheless, on the qualitative level, the predictions of the study
presented here are in good agreement with the observations. As a
matter of fact, (i)~in the regions with strong velocity shear,
Alfv\'enic correlations are reduced and a high level of density
perturbations are observed \citep{gr99}; (ii)~the spectrum of
density fluctuations in the inertial interval of the solar wind
fluctuations follows the Kolmogorov scaling $\rho^\prime \sim
k^{-5/3}$ \citep{mt90}. This agrees with the model developed by
\citet{mbm87}, where it has been shown that in the weakly
compressible limit, the density perturbations behave like a passive
scalar, i.e., they follow the scaling of the magnetic field
perturbations. The findings of the present paper supplement this
model and offer an efficient mechanism for the generation of the
density perturbations in the energy containing interval, which
afterwards are cascaded to the smaller scales.

Another interesting and not completely understood problem is the
spatial aspect of the velocity-shear-induced wave transformations in
the solar wind. If the largest values of the flow shearing rates
appear in the relatively narrow ``transition zones" between the slow
and the fast solar wind, one may wonder whether the
velocity-shear-induced mode conversions are confined to these
regions or whether they tend to occupy a wider volume of the wind
plasma!? Evidently only real-space numerical simulations, taking
into account physical characteristics of the solar wind, may answer
this question. Direct numerical simulations, performed in the
general context of SWTs, have shown \citep{bprr01} that when the
Alfv\'en waves get transformed into the fast magneto-sonic waves,
the latter tend to leave their ``birthplace" and propagate quite
rapidly due to their high velocities. We suppose that the
observational signature of the waves generated by means of this
mechanism would be their appearance throughout both the sheared and
the non-sheared wind volume with the maximal probability of their
detection in the vicinity of the strongly sheared transition
regions.

In the near future, we would like to take into account the effect of
the kinematic complexity \citep{MR99} and verify whether the
transition matrix method can lead to a quantitative description of
the SWTs not only for the solar wind, but also for solar jet-like
flows with a more complicated geometry and complicated kinematics.

\acknowledgments
Andria Rogava wishes to thank the \emph{Katholieke Universiteit
Leuven} (Leuven, Belgium) and the \emph{Abdus Salam International
Centre for Theoretical Physics }(Trieste, Italy) for supporting him,
in part, through a Senior Postdoctoral Fellowship and a Senior
Associate Membership Award, respectively. The research of Andria
Rogava and Grigol Gogoberidze was supported in part by the Georgian
National Science Foundation grant GNSF/ST06/4-096. The research of
Grigol Gogoberidze was supported in part by INTAS grant
06-1000017-9258. Andria Rogava and Grigol Gogoberidze are grateful
for the hospitality to the Abdus Salam International Centre for
Theoretical Physics, where a part of the work was done during their
visits to the ICTP, in 2005, as a Senior Associate and a Young
Collaborator, respectively. These results were obtained in the
framework of the projects GOA 2004/01 (K.U.Leuven), G.0304.07
(FWO-Vlaanderen) and C~90203 (ESA Prodex 8).

\newpage

Figure captions

FIG.~1: Transformation coefficient $ T_{fA}$ vs $\delta$.
Dash-dotted line and dashed line represent analytical expressions
(26) and (27), respectively. The solid line is obtained by the
numerical solution of Eqs.~(7)--(9).

FIG.~2: Transformation coefficient $T_{sA} $ vs $K_y$ for $R=0.005$
and $\beta=5$ and $\beta=200$.

\newpage

\begin{figure}[t]
\includegraphics[]{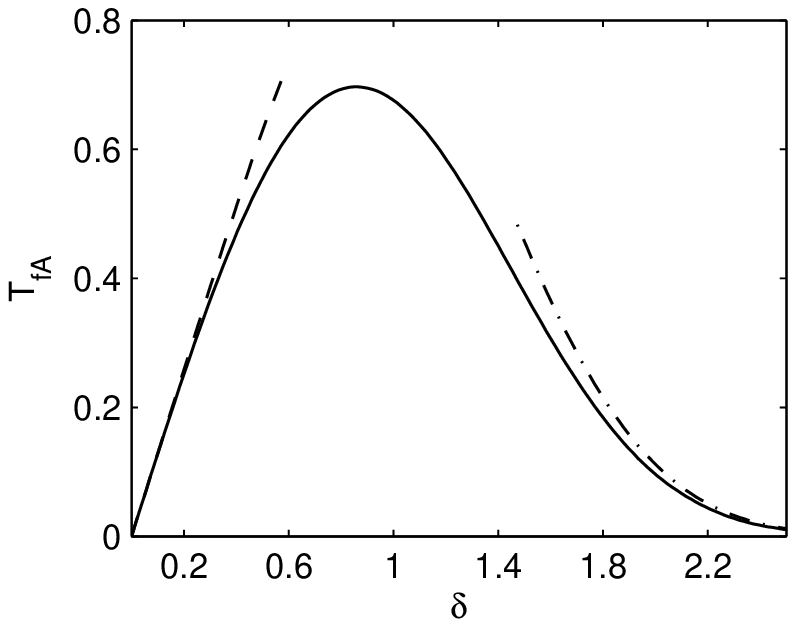}
\caption{\label{fig:fig1}}
\end{figure}

\begin{figure}[t]
\includegraphics[]{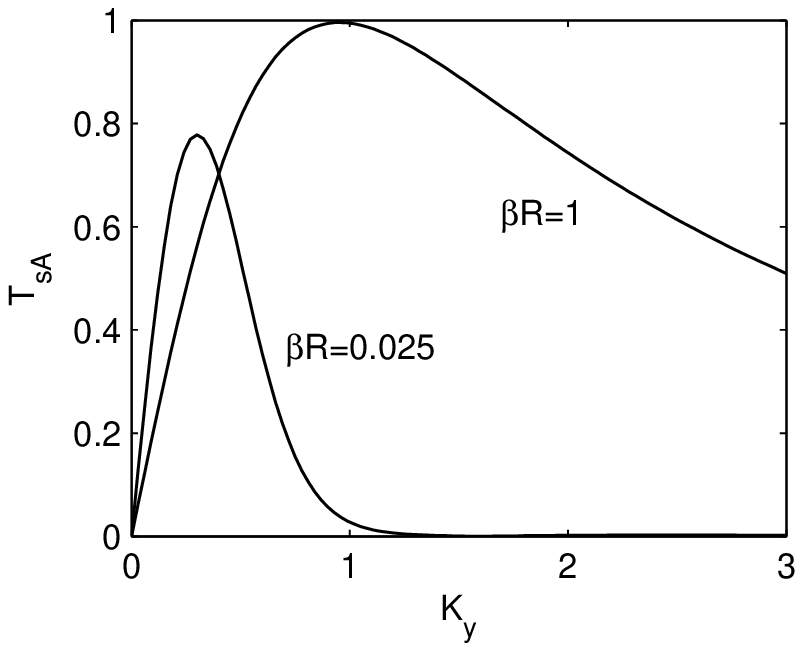}
\caption{\label{fig:fig2}}
\end{figure}

\end{document}